\documentclass[sigconf]{acmart}

\usepackage{gensymb}
\usepackage{booktabs}

\definecolor{brown}{rgb}{0.59, 0.29, 0.0}
\definecolor{darkgray}{rgb}{0.59, 0.59, 0.59}
\definecolor{tablegray}{gray}{.9}

\newcommand\rv[1]{#1}

\newcommand{\system}{Auptimize}

\usepackage[utf8]{inputenc}
\usepackage{diagbox}
\usepackage{colortbl}

\usepackage{color}
\usepackage{soul}

\usepackage{tabularx}

\usepackage{nameref}

\usepackage{xspace}
\usepackage{enumitem}
\usepackage{mathtools}
\usepackage{commath}
\DeclareMathOperator*{\argmin}{argmin}

\usepackage{amssymb}
\usepackage{pifont}

\newcommand{\customtilde}{{\raise.17ex\hbox{$\scriptstyle\sim$}}}

\newcommand{\etal}{et~al.\xspace}
\newcommand{\eg}{e.\,g.,\xspace}
\newcommand{\ie}{i.\,e.,\xspace}

\usepackage{xparse}

\sethlcolor{yellow}

\AtBeginDocument{%
  }

\copyrightyear{2024}
\acmYear{2024}
\setcopyright{rightsretained}
\acmConference[UIST '24]{The 37th Annual ACM Symposium on User Interface Software and Technology}{October 13--16, 2024}{Pittsburgh, PA, USA}
\acmBooktitle{The 37th Annual ACM Symposium on User Interface Software and Technology (UIST '24), October 13--16, 2024, Pittsburgh, PA, USA}
\acmDOI{10.1145/3654777.3676424}
\acmISBN{979-8-4007-0628-8/24/10}

\begin{document}
\title{Auptimize: Optimal Placement of Spatial Audio Cues for Extended Reality}

\author{Hyunsung Cho} 
\orcid{0000-0002-4521-2766}
\affiliation{%
  \institution{Carnegie Mellon University}
  \city{Pittsburgh}
  \state{PA}
  \country{USA}}
\email{hyunsung@cs.cmu.edu}

\author{Alexander Wang} 
\orcid{0009-0001-4353-4737}
\affiliation{%
  \institution{Carnegie Mellon University}
  \city{Pittsburgh}
  \state{PA}
  \country{USA}}
\email{aw4@andrew.cmu.edu}

\author{Divya Kartik} 
\orcid{0009-0005-5136-540X}
\affiliation{%
  \institution{Carnegie Mellon University}
  \city{Pittsburgh}
  \state{PA}
  \country{USA}}
\email{dkartik@andrew.cmu.edu}

\author{Emily Liying Xie} 
\orcid{0009-0005-3160-6579}
\affiliation{%
  \institution{Carnegie Mellon University}
  \city{Pittsburgh}
  \state{PA}
  \country{USA}}
\email{elx@andrew.cmu.edu}

\author{Yukang Yan}
\orcid{0000-0001-7515-3755}
\affiliation{%
  \institution{University of Rochester}
  \city{Rochester}
  \state{NY}
  \country{USA}}
\email{yanyukanglwy@gmail.com}

\author{David Lindlbauer}
\orcid{0000-0002-0809-9696}
\affiliation{%
  \institution{Carnegie Mellon University}
  \city{Pittsburgh}
  \state{PA}
  \country{USA}}
\email{davidlindlbauer@cmu.edu}

\renewcommand{\shortauthors}{Cho et al.}

\begin{abstract}
Spatial audio in Extended Reality (XR) provides users with better awareness of where virtual elements are placed, and efficiently guides them to events such as notifications, system alerts from different windows, or approaching avatars.
Humans, however, are inaccurate in localizing sound cues, especially with multiple sources due to limitations in human auditory perception such as angular discrimination error \rv{and} front-back confusion.
This decreases the efficiency of XR interfaces because users misidentify from which XR element a sound is coming.
To address this, we propose Auptimize, a novel computational approach for placing XR sound sources, which mitigates such localization errors \rv{by utilizing the ventriloquist effect.}
Auptimize disentangles the sound source locations from the visual elements and relocates the sound sources to optimal positions for unambiguous \rv{identification of} sound cues, avoiding errors due to inter-source proximity and front-back confusion.
Our evaluation shows that Auptimize \rv{decreases} spatial audio-based source identification errors compared to playing sound cues at the paired visual-sound locations.
We demonstrate the applicability of Auptimize for diverse spatial audio-based interactive XR scenarios.
\end{abstract}


\begin{teaserfigure}
    \vspace{-1em}
  \includegraphics[width=\textwidth]{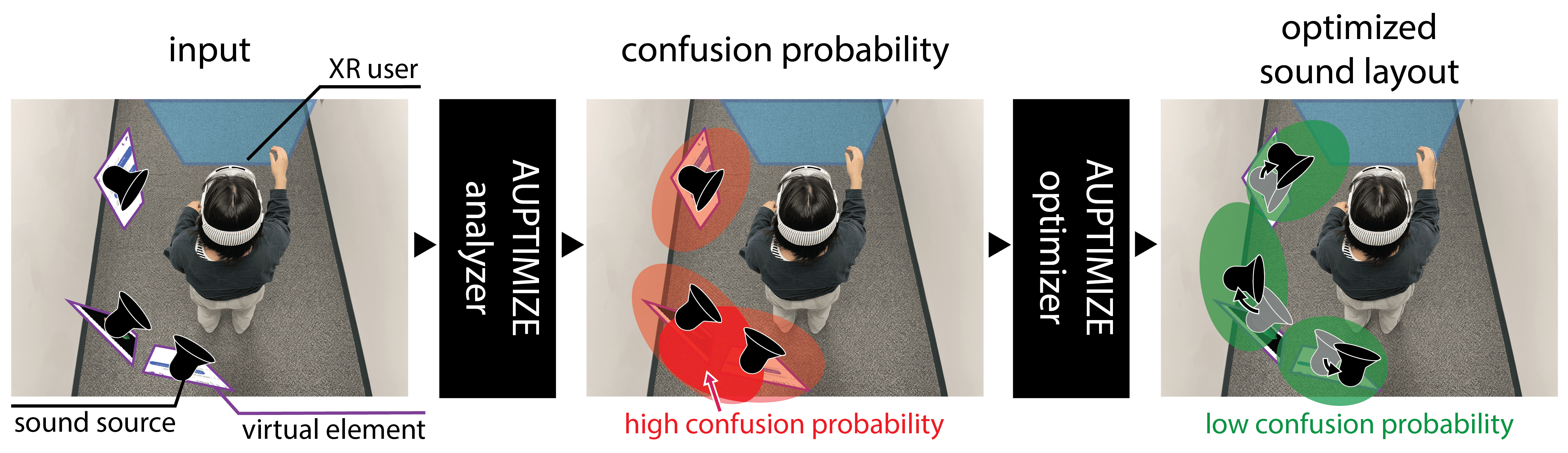}
  \caption{Auptimize takes the layout of Extended Reality (XR) elements as input (left), analyzes the \rv{probability} that the user misidentifies from which element a spatial sound is coming from (middle), then relocates the sound sources to the optimized locations which are disentangled from their visual counterparts to minimize the confusion probability (right).}
  \Description{Figure showing a user wearing an Augmented Reality headset from bird-eye view. Virtual elements are behind them. The figure illustrates that audio notifications do not correctly identify target objects due to perceptual limitations. Auptimize relocates the sound sources to the optimized locations which are disentangled from their visual counterparts to minimize the confusion probability.}
  \label{fig:teaser}
\end{teaserfigure}


\maketitle

\section{Introduction}
Auditory cues for Extended Reality (XR) interfaces enable users to interact with spatial user interfaces efficiently, and help guide them through the vast user interaction space~\cite{franinovic2013sonic}.
While XR devices can display interface elements at arbitrary locations, including behind users, they typically have a limited field of view.
This makes spatial audio cues especially helpful for tasks such as target acquisition of out-of-view objects, where visual cues may not be sufficient for navigating the user interfaces (UI) with gaze, controllers, or hand gestures.
For example, in XR, spatialized audio notifications deliver information promptly to users while hinting where the users needs to orient to for finding the relevant information.
This becomes more important when XR interfaces are increasingly distributed to users' periphery~\cite{lu2020glanceable} and integrated into their ambient surroundings~\cite{han2023blendmr}.

Despite its benefits, spatial audio cues for XR interaction are prone to errors because humans are inherently inaccurate in localizing sound sources in space due to perceptual limitations of the human auditory system. 
First, human ears experience \textit{localization blur} where they cannot accurately pinpoint spatial sound sources, especially compared to vision. 
Secondly, \textit{front-back confusion}, or more broadly the \textit{cone of confusion}, leads to inaccuracies in identifying whether a sound source is in the front or back of the user.
This is because sound sources that are in the front or the back produce similar interaural time differences (ITD) and interaural level differences (ILD), which are crucial for sound localization. 

In this work, we address this challenge and contribute \system{}, a computational approach to mitigate such confusion.
Our approach displaces spatial audio cues to optimal locations while preserving the location of visual elements.
This enables users to disambiguate spatial audio cues more accurately.
\autoref{fig:teaser} illustrates an example: an XR user has opened multiple conversation windows and a feed of stories around them. 
The user performs a primary task, the other UI elements are in their periphery or behind them, specifically the messenger interfaces.
When a new message arrives in one of the conversations, it makes a spatial auditory notification played from the location of the conversation window (\ie behind users). 
However, because the conversation window is spatially close to another conversation or feed, the user experiences confusion in identifying which one of the three windows actually has an update.
They thus have to inspect each UI element individually to identify the notification sound sources. 
This leads to a delayed response, and increases the chance that users miss out on important information.

To enable more accurate localization of spatial audio, we contribute \system{} with its two key components, the \textit{\system{} analyzer} and the \textit{\system{} optimizer}. 
Given a layout of virtual elements in an XR interface, the \textit{analyzer} predicts inaccuracies and confusions in the source identification of spatial XR audio cues (\autoref{fig:teaser} middle).
We achieve this by contributing a novel computational model that enables identifying such localisation errors.
The model is based on a data collection ($n = 15$) where we measured auditory localization errors and confusion patterns with spatial audio cues in XR. 
As a second step, the optimizer first disentangles the audio location from the (visual) location of the XR element, and then moves the spatial audio cue in real-time to a location that lowers the probability of confusion among virtual elements~(\autoref{fig:teaser} right).
We leverage aforementioned computational model and use integer programming to solve the problem of assigning sound sources to optimal locations efficiently and in real-time.

We evaluate our approach in a user study ($n = 12$) comparing it to a spatial audio created from a generic head-related transfer function (HRTF) and dynamic audio cues.
\system{} outperforms both in disambiguation performance.
\rv{This is a crucial first step towards finding the optimal combination of multiple modalities to deliver information in spatial interfaces}
We demonstrate the potential of our approach with a set of example applications where using spatial audio cues in XR benefit from enhanced accuracy with \system{}.




\section{Background and Related work}
\begin{figure}[ht]
    \centering
    \includegraphics[width=0.7\linewidth]{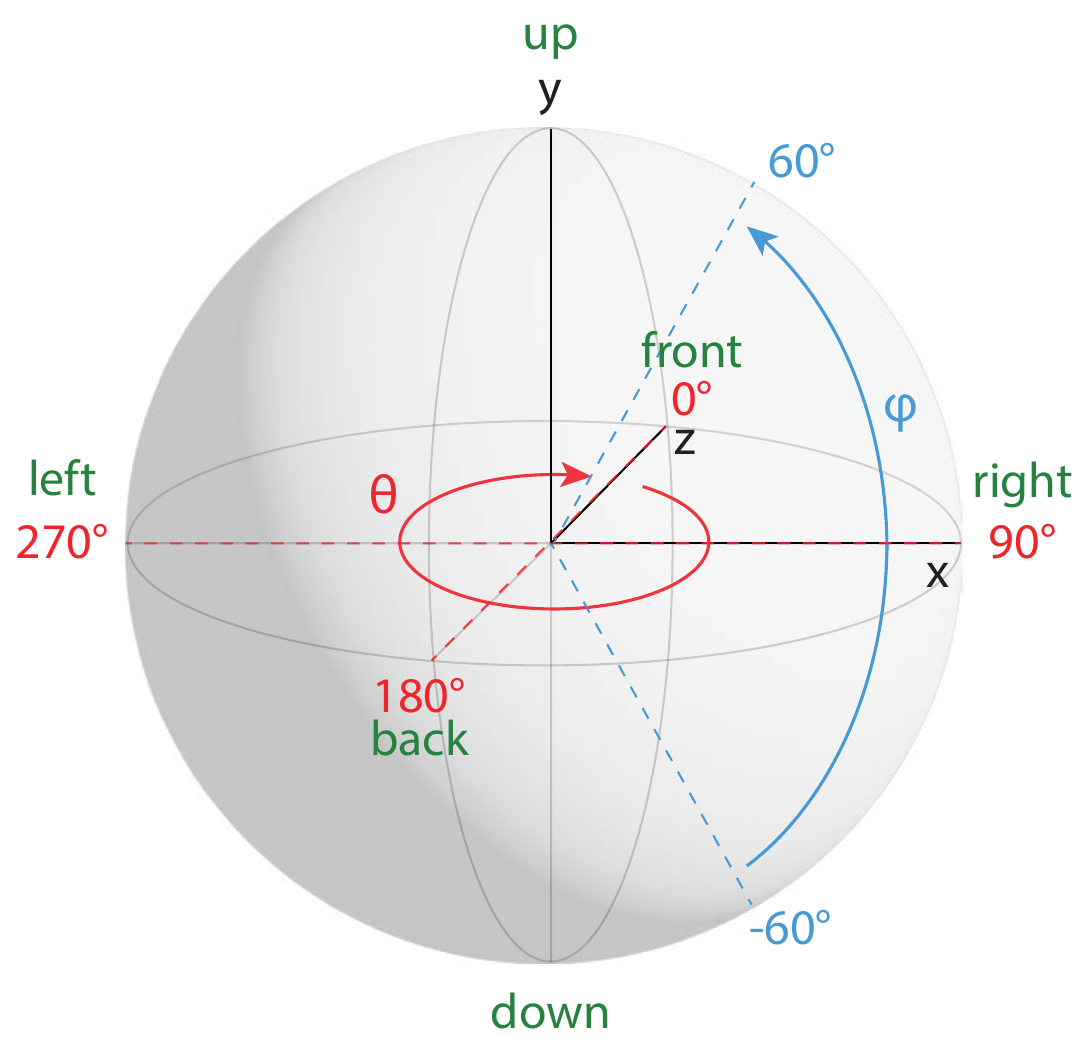}
    \caption{Spherical and rectangular coordinate systems used in this work.}
    \Description{Illustration of a 3D coordinate system.}
    \label{fig:coordinate_system}
\end{figure}

Our method builds on the understanding of how humans localize a sound and is designed to mitigate perceptual errors involved in the sound localization.
We provide an overview of the backgrounds in human sound localization process, errors, and perceptual illusions, necessary to understand how \system{} works.
Knowledgeable readers can skip this part as it contains well-known foundational knowledge.
In the remainder of the paper, we define the auditory space relative to the observer, as illustrated in Figure~\ref{fig:coordinate_system}. 
The coordinate system we use throughout the paper, `front', `right', `back', and `left' referring to the relative direction of the observer.
The angle $\theta$ denotes an angle in the azimuth or the horizontal plane of auditory space.
An elevation angle in the vertical plane is denoted by $\phi$. 

\subsection{How Humans Localize Sound} \label{sec:background_localization}
Humans localize sound in the horizontal plane using two primary binaural cues: interaural time difference (ITD) and interaural level difference (ILD).
ITD refers to the difference between the \textit{times} that sounds reach the two ears. 
By comparing the time of the sound signal's arrival at each ear, the human auditory system translates the interaural time difference into the azimuth angle of the sound.
ILD is the difference in the sound pressure level reaching the two ears, also known as interaural intensity difference (IID). 
This difference occurs because the human head casts an \textit{acoustic shadow}, reducing the sound level for the far ear, especially for high frequency sounds (>1500Hz) that have short wavelengths.

In a controlled natural environment, azimuth localization error varies from $\pm$3.6\degree~ when the noise is in front, $\pm$10\degree~ to the sides, and $\pm$5.5\degree~ for sounds coming from behind the listener, when sound sources were presented through loudspeakers~\cite{blauert1997spatial}.
The inaccuracy is also referred to as \textbf{localization blur}.

\begin{figure}[t]
    \centering
    \includegraphics[width=\columnwidth]{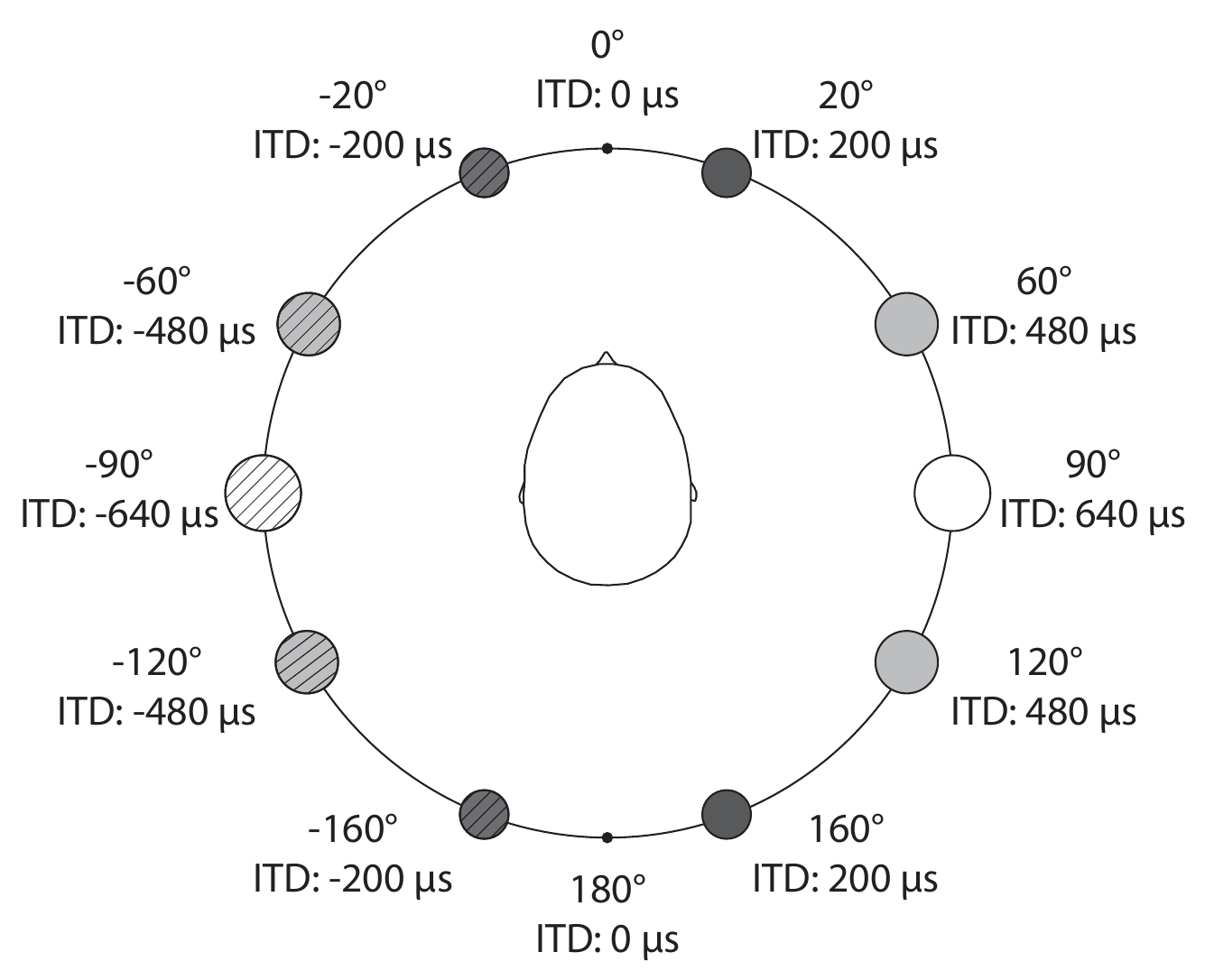}
    \caption{Interaural time differences (ITD) per azimuth angle.}
    \Description{Illustration of how angles relative to user's head are defined. 0 degree is when users look straight, 90 degrees to their right.}
    \label{fig:itd}
\end{figure}

\begin{figure}[t]
    \centering
    \includegraphics[width=0.4\columnwidth]{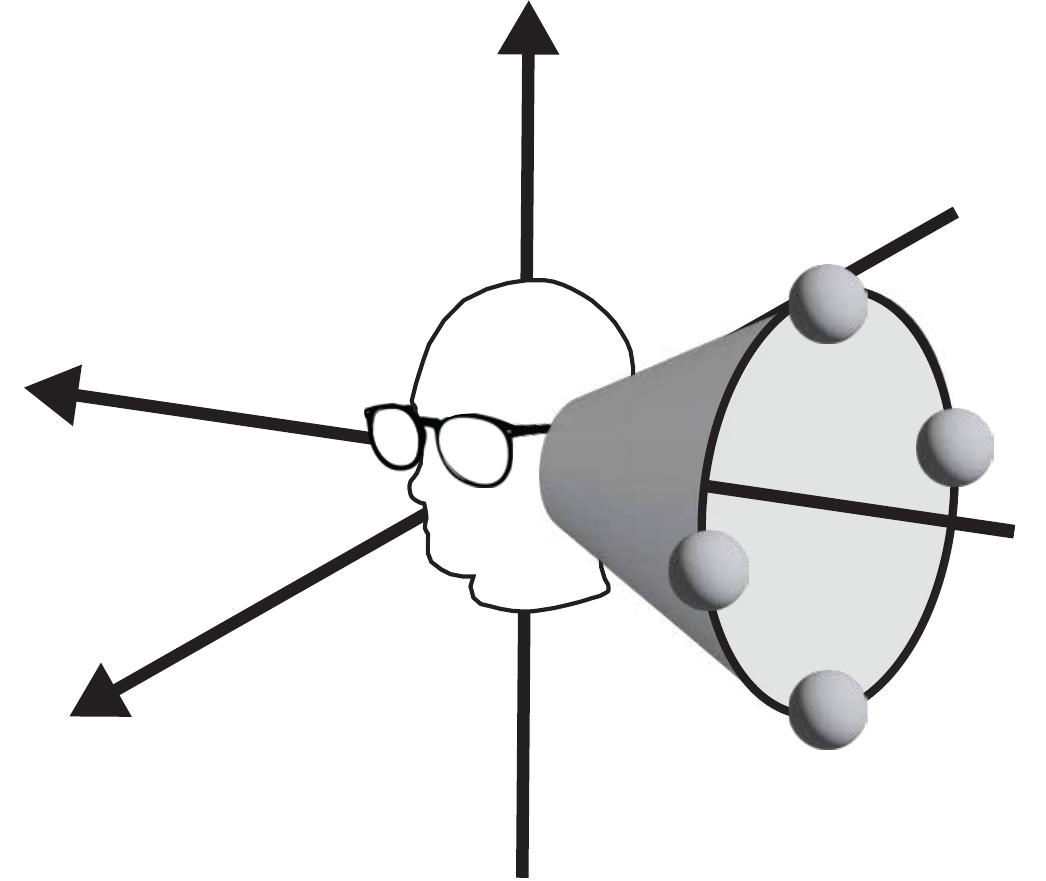}
    \caption{Cone of confusion, where the points at the cross-sections of the cone have the same interaural time differences.}
    \Description{Illustration of the cone of confusion. A cone originates in users' ear. Audio sources on this cone are frequently confused with each other.}
    \label{fig:cone_of_confusion}
    \vspace{-1em}
\end{figure}

Azimuth localization based on ITD and ILD has several limitations. 
As illustrated in Figure~\ref{fig:itd}, sound sources at the front and back of the head have the same ITD, resulting in \textbf{front-back confusion}.
If we extend this phenomenon to the 3D space considering elevation, this forms a \textbf{cone of confusion}~\cite{wallach1940role}, as shown in Figure~\ref{fig:cone_of_confusion}, where points on the cone's cross-sections are equidistant from the left and right ears. 
\textit{In other words, sounds played at different locations on the cone of confusion are hardly distinguishable as they produce the same ITDs and similar ILDs.}
Humans resolve the ambiguities in elevations by rotating heads and altering the binaural cues with a concurrent shift. 
Tilting the head also resets the problem of elevation localization to azimuth localization.
For this reason, we scope our work to azimuth localization based on spatial audio.

%
%
%
%
According to Stevens and Newman~\cite{stevens1934localization}, when localizing pure tones, the greatest uncertainty occurs at 1500 Hz.
Localizing natural sounds, often consisting of complex waveforms or wideband noises, is easier than that of pure tones because the ITDs provide more cues. 
For brief sounds, onsets and offsets are critical to provide unambiguous cues for localization.
Most auditory cues used in computer systems (\eg notification sounds or system alerts) are brief, complex sounds with clear onsets, which are more favorable for localization.
In our work, we employ these brief sounds of complex waveforms to test our method in a realistic XR usage setting.
We reveal the limitations in localizing even these favorable sounds in the current spatial audio system and propose a method to improve disambiguation of sound sources. 
In addition to ITDs and ILDs, the \textit{head-related transfer function (HRTF)} also affects sound localization.
HRTF refers to the changes in frequency intensity that occur as sound waves interact with the anatomical features of a listener, \eg pinnae or head. 
These anatomical elements sculpt the sound through reflection, diffraction, and absorption, modifying the sound waves' paths and thereby influencing the frequency intensities that are perceived. 

\subsection{Ventriloquist Effect}
In auditory-visual interactions in space, vision can dominate auditory spatial localization~\cite{harris1965perceptual}. 
This auditory illusion is often referred to as the \textit{ventriloquist effect} where the origin of a sound is incorrectly perceived as coming from a visible entity, even though the actual sound source is unseen and located elsewhere.
Kytö~\etal\cite{kyto2015ventriloquist} studied the ventriloquist effect in Augmented Reality (AR), discovering that it is effective when the spatial disparity between an audio source and a visual source is within 30\degree~ of azimuth angle. 
Gorzel~\etal\cite{gorzel2012distance} confirmed the effect for distance perception.
This effect complicates sound localization further, particular for multiple virtual elements distributed around a user.
For example, due to front-back confusion, if virtual elements are placed at the front and back of a user, even though the element at the back of the user emits a sound, the user might perceive it coming from the visual element in the front as it is currently in their field of view.

On the flip side, the ventriloquist effect serves as the main motivation of our work, since we observe that \textit{the actual location of an auditory cue does not have to perfectly align with its visual element} while they can still be perceived as one complete whole. 
\system{} leverages this effect and finds the optimal locations of spatial audio cues and displace them while preserving the original visual locations.

\subsection{Localization in Digital Systems}
Spatializing audio in digital systems, including virtual reality, focuses on simulating the binaural cues of ITDs, ILDs, and HRTFs described in Section~\ref{sec:background_localization}.
Audio spatializers model the physics of the virtual environment and compute the expected ITDs and ILDs based on the model.
HRTF is a more personal component because this depends on the structures of pinnae and head that are unique to each user.
Audio spatializers in digital systems involve a generic HRTF model or, more recently, personalized HRTF models based on a visual capture and reconstruction of one's ear and head.
Although more accurate, personalization involves an arduous process of calibrations to the unique anatomical features of the user.
Berger et al.~\cite{berger2018generic} suggest that due to the high degree of cross-modal plasticity in cortical sensory processing, like the ventriloquist effect, generic (non-individualized) HRTFs may suffice for auditory spatial localization in VR.

Valzolgher et al.~\cite{valzolgher2020impact} conducted a localization study in virtual reality where real free-field stimuli of 3-second bursts of white noise were played through hidden loudspeakers. 
They tested a combination of 4 different azimuths ($\pm$45\degree~ and  $\pm$ 135\degree), a single elevation at the ear level, and 2 distances (35cm and 55cm). 
The results show that azimuth error was around 11.5\degree~with worse performance for the back (M=13.5\degree) than front (M=9.4\degree) and elevation error around 13.5\degree.
Wenzel et al.~\cite{wenzel1993localization} conducted a study to measure localization error when sounds stimuli (spectrally shaped bursts of Gaussian noise) are transduced by loudspeakers or headphones. 
For the headphone conditions, each sound was digitally processed, simulating the direction, outer ear characteristics (HRTFs) measured for a representative subject, following the synthesis procedure by Wightman and Kistler~\cite{wightman1989aheadphone, wightman1989bheadphone}.
The results showed localization error of around 16 to 23 degrees azimuth around the actual sound source.

The performance of localization in virtual systems heavily relies on the performance of underlying spatializer. 
In this work, we conducted a sound localization study (Section~\ref{sec:data_collection}) using the HRTF spatial audio from the Oculus Audio SDK version 47.0, which was the state-of-the-art at the study time.
We contribute the findings on localization error of sound stimuli processed with non-individualized HRTF in a virtual environment, delivered through headphones, and we use these errors as the basis to develop \system{}.

\subsection{Spatial Audio for Interactive Systems}
Spatial audio has been used in human-computer interaction to help with information presentation~\cite{chang2022omniscribe}, directing user attention~\cite{peng2018speechbubbles, chang2024sound}, and creating richer spatial experiences such as spatial videos~\cite{bala2019elephant}, mid-air interaction~\cite{Muller2014-ot}, gaming~\cite{broderick2018importance}, music listening~\cite{lim2021spatial}, ambient interfaces~\cite{Ishii2001-ws}, or conferencing~\cite{ji2022vrbubble}.
\citeauthor{Iravantchi20}\cite{Iravantchi20} leverage the ventriloquism effect to spatialize audio towards physical objects.

In conversations, spatial audio increases the perception of interactivity~\cite{nowak2023hear}, while in a desktop environment, it can increase memory and perceive comprehension~\cite{baldis2001effects}.
While \citeauthor{inkpen2010exploring}~\cite{inkpen2010exploring} did not find significant benefits of replacing mono audio with spatialized audio, \citeauthor{zhong2022binaural}~\cite{zhong2022binaural} show that binaural audio is helpful in hybrid formats and \citeauthor{hyrkas2023spatialized}~\cite{hyrkas2023spatialized} highlight benefits for remote attendees in video calls.
Finally, spatial audio improves non-visual navigation, 
 as discussed by \citeauthor{Loomis1993}~\cite{Loomis1993}.
We leverage spatial audio for its ability to give localized cues.






\paragraph{Spatial Audio in XR}
Spatial audio is particularly useful in XR settings where virtual elements can be presented in the 3D space around users~\cite{cho2024minexr}.
Using a spatial audio cue is known to reduce search time of a virtual element around a user by 35\% compared to without using a cue~\cite{billinghurst1998evaluation}.
Auditory cues reduce the time to locate out-of-view objects in head-mounted AR although the accuracy was not affected~\cite{binetti2021using}. 
In spatial conferencing, spatial cues improved the performance in speaker discrimination significantly with higher rating~\cite{billinghurst1998wearable}.
\citeauthor{dong2016towards}\cite{dong2016towards} show that spatial audio improves navigation performance, whereas \citeauthor{huang2021using}~\cite{huang2021using} show that it is an effective mechanism to combat distance compression.
%
%
Work by \citeauthor{buck2022effect}~\cite{buck2022effect}, \citeauthor{mcmullen2014potentials}~\cite{mcmullen2014potentials} and \citeauthor{poeschl2013integration}~\cite{poeschl2013integration} shows that spatial audio positively influences aspects such as preservation of personal space, immersion, and presence, respectively.
It is unclear, though how large this effect is, since work by \citeauthor{hendrix1995presence}~\cite{hendrix1995presence} showed that visual features such as field of view or head tracking might have even more influence on presence.
For 360\degree{} videos, spatial audio nudged users towards more exploration~\cite{hirway2022spatial}, and changed their focus towards sound-emitting regions~\cite{hirway2020qoe, chao2020audio, singla2023saliency}.
Our work contributes a new method to make sound localization more accurate, which would be beneficial for many of above approaches.
We refer readers to surveys and overviews by \citeauthor{cohen2015special}~\cite{cohen2015special}, \citeauthor{chaurasia2023challenges}~\cite{chaurasia2023challenges}, and \citeauthor{begault20003}~\cite{begault20003} for more in-depth discussions.

\paragraph{Audio Cues}
A dynamic audio cue moving in the direction of the visual target position further reduces onset time for a target acquisition, compared to a static audio cue, by directing user's attention effectively towards the target~\cite{barde2016attention}.
We use this dynamic audio cue as one baseline in our comparative evaluation (Section~\ref{sec:evaluation}).

While the concept of distancing the audio cue from its visual counterpart has been used to experiment with the ventriloquist effect in previous studies, it has been underexplored how the distancing can be utilized to improve auditory source disambiguation.
Recent work explored adding offsets to spatial audio cues to compensate for the perceptual bias induced by head-body rotations~\cite{bernal2024modeling}.
However, it does not address the localization blur and cone of confusion which are present even when a user keeps their head and body still.
Our work proposes a novel computational method, \system{}, to compensate for the localization blur and cone of confusion that hinder auditory source identification by introducing optimal offsets to the auditory cues to minimize confusion.

\section{Data Collection} \label{sec:data_collection}
We collected data to \rv{quantify} auditory localization errors in XR and confusion patterns when spatial audio cues are played through the XR audio spatializer at different angles. 
\rv{While prior work provides error bounds at certain angles, we needed to collect data from various angles to build the optimizer.}
This data serves as the ground truth data for our optimization algorithm to mitigate the confusions in audio source identification.
The study was conducted using an Meta Quest 2 VR headset. The data collection program was developed using Unity version 2021.3.9 and HRTF spatial audio from the Oculus Audio SDK version 47.0.
The study involved 15 participants, aged between 20 and 30 years (5 male, 10 female; $M = 24.4$, $SD = 2.87$), all students and staff from a local University.
\rv{Our data is available at \textit{\url{https://augmented-perception.org/publications/2024-auptimize.html}}.}

\subsection{Study Design and Procedure}
\begin{figure}[t]
    \centering
    \includegraphics[width=\columnwidth]{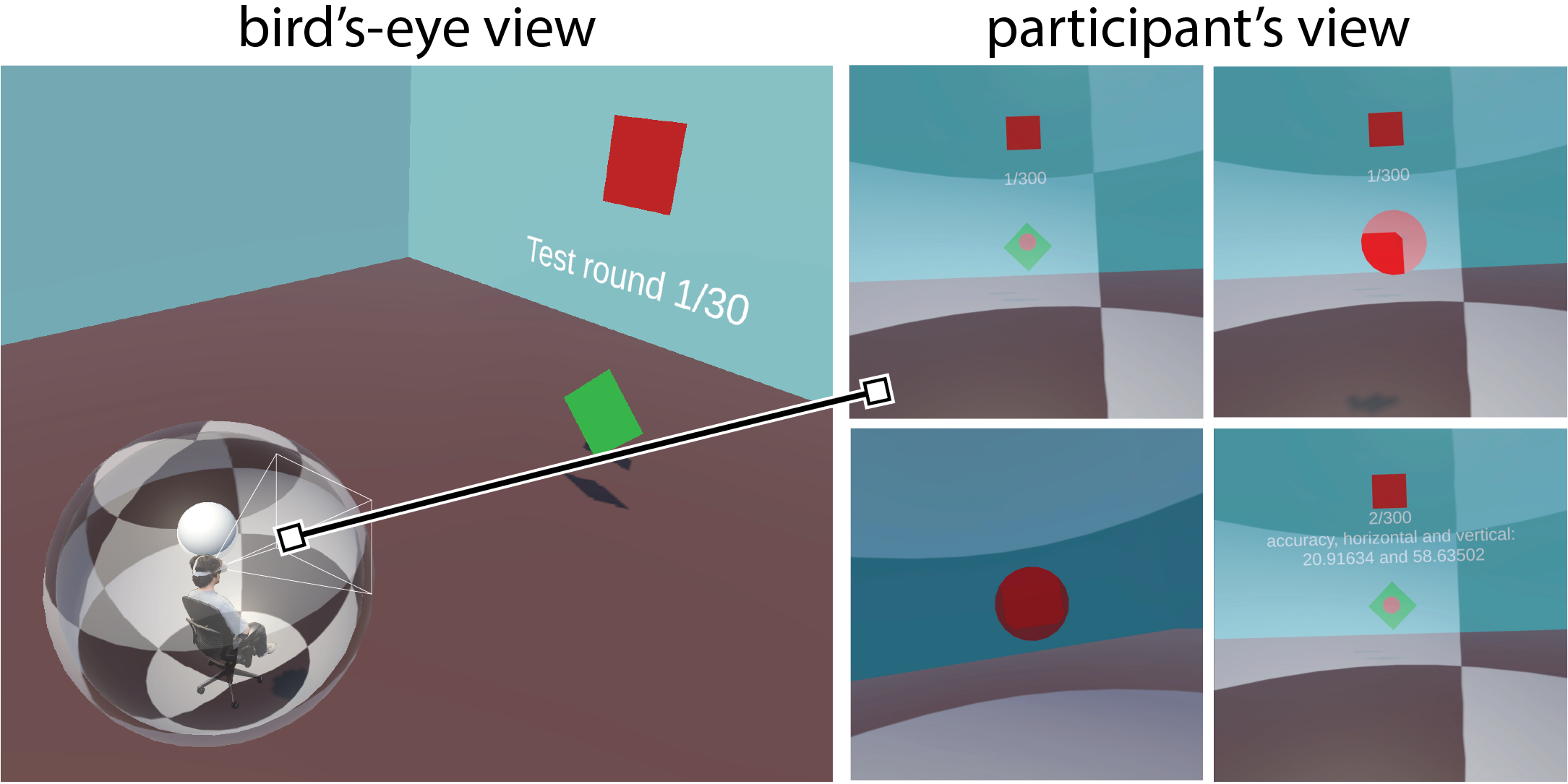}
    \caption{Data collection environment of spatial audio localization. Participants were seated, and asked to point at the location of sound sources on the semi-transparent sphere with checker board.}
    \Description{Data collection environment of spatial audio localization. Participants were seated, and asked to point at the location of sound sources on the semi-transparent sphere with checker board.}
    \label{fig:data_collection_setup}
\end{figure}

Participants were placed in a Virtual Reality (VR) environment, seated while wearing a head-mounted display, as illustrated in \autoref{fig:data_collection_setup}.
We chose a VR \rv{paradigm to control the environment of the study}.
In each trial, participant were asked to face forward, and start the next trial by pressing a button with the VR controller. 
For each trial, a spatialized 3-second white noise was played.
Participants were asked to rotate their head towards the direction they believe the sound came from, and confirm the location with by pressing the controller button.
The location did not have a visual representation.
For each trial, they received information about the distance from their chosen location to the sound source location.
Each participant completed 30 training trials, and 300 trials for data collection.
The azimuth and elevation angles of the sound sources tested in the 300 trials were randomly generated following a uniform distribution within the range of $0\degree\leq\theta<360\degree$ for azimuth and $-60\degree\leq\phi<60\degree$ for elevation. 
The sound sources were placed in the assigned direction \rv{at} a distance of 1.5 meters.

\subsection{Results} 
\begin{figure*}[t]
    \centering
    \vspace{-0.5em}
    \includegraphics[width=\linewidth]{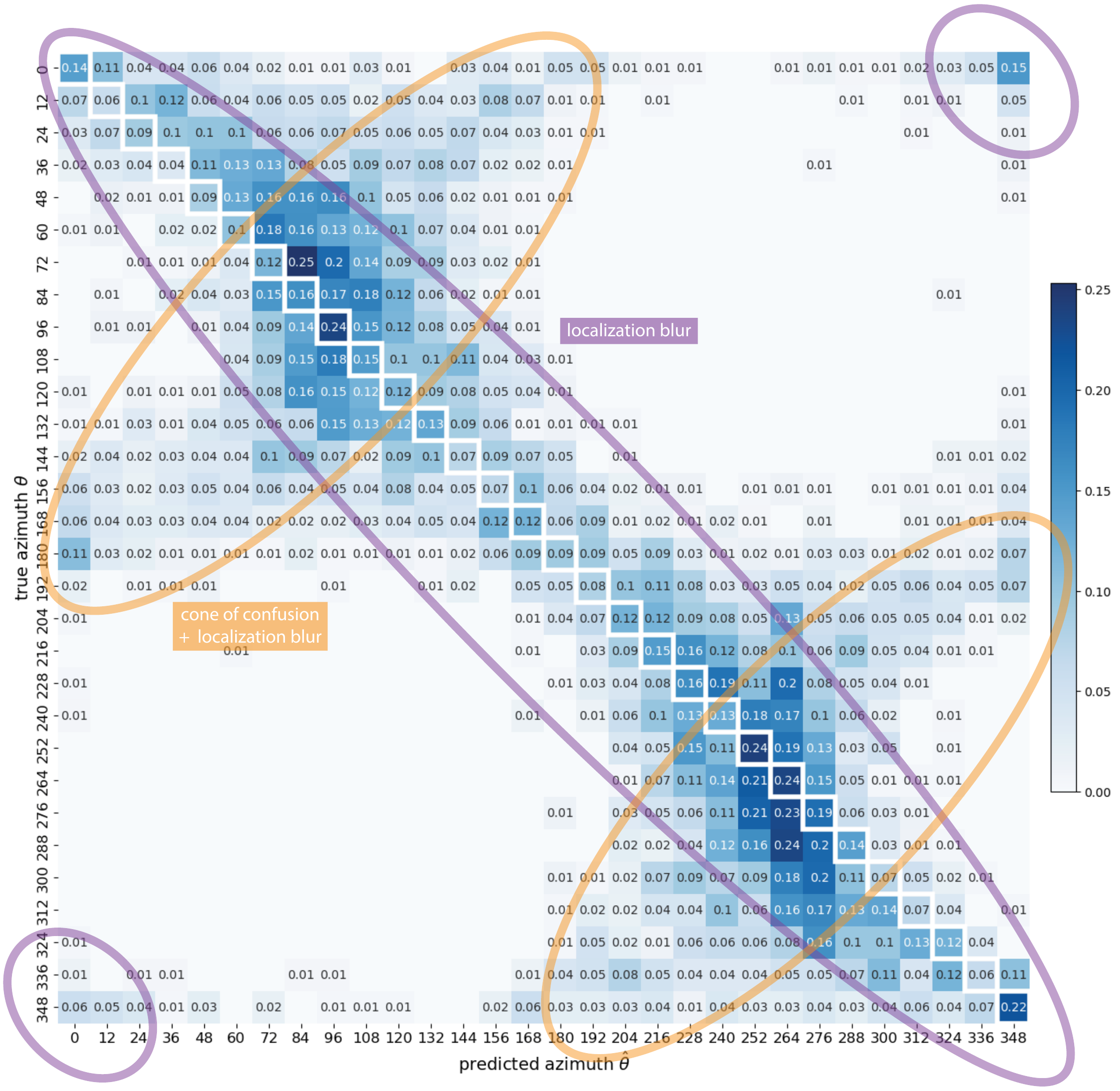}
    \vspace{-0.5em}
    \caption{Confusion matrix of azimuth localization in the data collection study, discretized into 30 bins of size 12\degree. Each cell represents the probability of a sound played at a true azimuth angle $\theta$ to be perceived as coming from the predicted azimuth angle $\hat{\theta}$. Cells in purple ellipses indicate the effects of localization blur, and cells in orange ellipses exhibit the effects of localization blur around the cone of confusion.}   
    \Description{Confusion matrix of azimuth localization in the data collection study, discretized into 30 bins of size 12 degree. Each cell represents the probability of a sound played at a true azimuth angle to be perceived as coming from the predicted azimuth angle.}
    \label{fig:confusion_matrix}
\end{figure*}

We collected a total of 4,500 data points.
Our analysis focuses on the azimuth localization errors, which are the main target of \system{}'s optimization.
In \autoref{fig:confusion_matrix}, we show the discretized results of participants' prediction of a sound's azimuth angle $\hat{\theta}$ (x-axis) coming from the true azimuth angle $\theta$ (y-axis) into bins of size 12\degree{}.
As illustrated in \autoref{fig:coordinate_system}, participants' front is 0\degree{} or 360\degree, right is 90\degree, back is 180\degree, and left is 270\degree.

\begin{table*}[h]
\centering
\caption{Summary of localization errors for front, right, back, and left regions. Circular error refers to the spherical angular error in azimuth $\theta$. Adjusted error refers to the spherical angular error after compensating for the front-back and up-down confusion, and cone of confusion effect is the difference between the two.}
\begin{tabular}{lccc}
\toprule
\textbf{Region} & \textbf{Circular error mean (SD)} & \textbf{Adjusted error mean (SD)} & \textbf{Cone-of-confusion effect mean (SD)} \\
\midrule
Front (-36\degree, 36\degree) & 57.83\degree{} (51.14\degree{}) & 27.03\degree{} (21.08\degree{}) & 30.79\degree{} (30.06\degree{}) \\
Right (36\degree, 144\degree) & 32.60\degree{} (26.63\degree{}) & 19.37\degree{} (14.87\degree{}) & 13.23\degree{} (11.76\degree{}) \\
Back (144\degree, 216\degree) & 62.88\degree{} (54.65\degree{}) & 28.40\degree{} (21.41\degree{}) & 34.48\degree{} (33.24\degree{}) \\
Left (216\degree, 324\degree) & 28.37\degree{} (24.04\degree{}) & 16.97\degree{} (13.24\degree{}) & 11.40\degree{} (10.80\degree{}) \\
\midrule
All & 42.45\degree{} (41.53\degree{}) & 22.00\degree{} (17.96\degree{}) & 20.45\degree{} (23.57\degree{}) \\
\bottomrule
\vspace{-1em}
\end{tabular}
\label{tab:localization_errors}
\end{table*}

\subsubsection{Localization Blur}
Cells on the diagonal line with white borders in \autoref{fig:confusion_matrix} present the ideal cases when accurate predictions are made, \ie users correctly located the sound.
Confusion in nearby cells around the white borders (circled in purple) represent the localization blur due to the perceptual inaccuracy in the azimuth localization. 
This is especially prominent in the right (around 36\degree{}-144\degree{}) and left (216\degree{}-324\degree{}) regions, where the predictions for front (324\degree{}-36\degree{}) and back (144\degree{}-216\degree{}) regions are spread out more widely across (fewer empty boxes which refer to zero occurrence of prediction).
Audio cues played at these side regions are rarely confused with the opposite side because the ITDs and ILDs are drastically different, as illustrated in \autoref{fig:itd}.

\autoref{tab:localization_errors} summarizes the localization errors in azimuth $\theta$ for front, right, back, and left regions.
The circular error represents the average spherical angular error in each dimension.
The adjusted error is a metric of spherical angular error which accounts for the front-back and up-down confusion~\cite{wenzel1993localization}.
This metric is computed by comparing the spherical angular error with the true angle of the audio stimulus and its flipped versions along the front/back axis and the up/down axis, and then taking the minimum error among all versions.
Therefore, the adjusted error only represents the effects of localization blur, which shows better precision in the sides (left and right) than in the front and back regions.

\subsubsection{Cone of Confusion}
The cells on the opposite side of each region in the front/back axis in \autoref{fig:confusion_matrix} fall on the cone of confusion.
The matrix illustrates the cone-of-confusion effect in combination with localization blur, where we observe distribution around two diagonal regions circled in orange.

The adjusted error in \autoref{tab:localization_errors} removes the effects of cone of confusion (front-back and up-down confusions). 
Thus, the differences between circular error and adjusted error can be considered as the effects of cone of confusion.
The cone of confusion effect appears greater in the front and back region than in the sides.


\subsubsection{Optimal Sound (Dis)placement} \label{sec:optimal_sound_displacement}
The confusion matrix reveals that the optimal location to place a single audio cue is not always same as the visual location.
If the optimal location is same as the visual location, the highest value in each row of the matrix must lie on the diagonal. 
However, for only 4 out of 30 bins (13.3\%), the optimal location corresponds to the visual location.
In other words, for 86.7\% of the cases, \textbf{displacing the audio cue's source location to a location that is different from the visual location improves localization}. This confirms our hypothesis that displacing the sound location from the visual location of a virtual element can guide a user's attention closer to the actual visual location, compared to same visual and sound locations.
We leverage the data from this model for our system.

\subsubsection{\rv{Aggregated Probability vs. Individual Users}}
\rv{We ran statistical tests to confirm that individual confusion patterns show no significant difference from an aggregated model (\autoref{fig:confusion_matrix}). 
All matrices showed small variations in Euclidean distance and strong Pearson correlations. 
Chi-Square and ANOVA tests yielded $p$-values of 1.0, indicating that differences are likely due to noise rather than systematic.
Therefore, we use the aggregated probability to build a generic version of \system{}.
}
\section{\system{}}
\begin{figure*}[t]
    \centering
    \includegraphics[width=\linewidth]{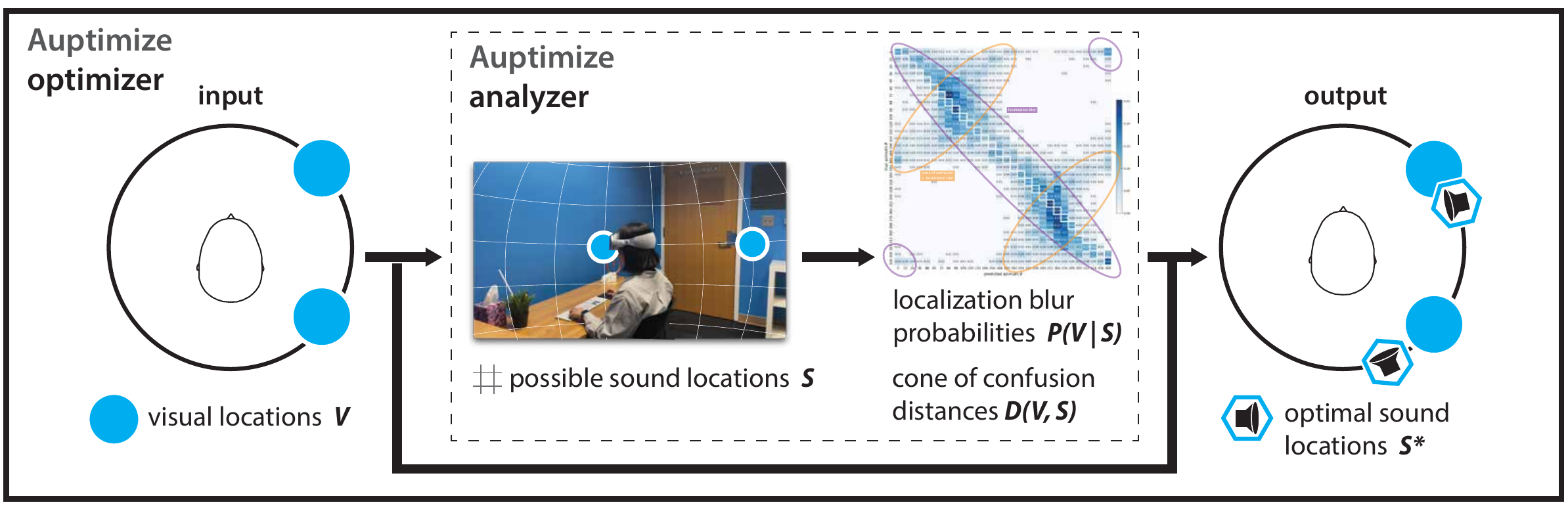}
    \caption{\system{} workflow: \system{}~takes the visual locations of an XR layout as input. Given each possible set of auditory locations, \system{} analyzer predicts the localization blur and cone of confusion. Based on the predictions, \system{} optimizer finds the optimal placement of the spatial audio cues.}
    \Description{Illustration of the Auptimize workflow: Auptimize takes the visual locations of an XR layout as input. Given each possible set of auditory locations, Auptimize analyzer predicts the localization blur and cone of confusion. Based on the predictions, Auptimize optimizer finds the optimal placement of the spatial audio cues.}
    \label{fig:system_workflow}
\end{figure*}

We contribute an optimization-based method to find the optimal placement of spatial audio cues to minimize localization error.
\system{} consists of two components: analyzer and optimizer.
The \system{} analyzer models the predicted localization blur and cone of confusion given a layout of virtual elements, and the \system{} optimizer uses these predicted confusion values to find the optimal placement of spatial audio cues.
The input for our system are visual locations given by an XR layout, the output are modified locations from which audio cues should be played from.
The workflow of \system{} is illustrated in \autoref{fig:system_workflow}.

\subsection{\system{} Analyzer}
We define a layout of XR elements as consisting of $n$ virtual elements, $E=(e_1, \ldots, e_n)$.
Each virtual element $e_i$ has a visual component located at its location $v_i$ and a sound component located at $s_i$.
The set of visual locations is denoted as $V=(v_1, \ldots, v_n)$ and sound locations as $S=(s_1, \ldots, s_n)$.
In this work, by ``locations'' we refer to the azimuth angle $\theta$ of the spherical coordinate as we fix the radius and preserve the original elevation angle $\phi$ in the optimization.
\rv{We varied elevation for rich data collection, but \system{} focuses on azimuth localization that uses binaural cues because elevation relies on monaural, spectral cues. 
Our approach modifies binaural cues via displacement, affecting azimuth localization. 
}

The \system{} analyzer takes $V$ and $S$ as input and outputs two metrics quantifying the expected localization blur and cone of confusion based on the data we collected (Section~\ref{sec:data_collection}).


\subsubsection{Localization Blur}
In the data collection, an audio cue was played at location $s$, and participants made a prediction $u(s)$ for its source location.
We define this prediction as a mapping $u: S \to V$, where $u(s)$ is the perceived location for $s$.
Aggregating all predictions constructs a probability distribution $P(V|S)$.
This corresponds to the rows of \autoref{fig:confusion_matrix} where $S$ is the true azimuth $\theta$ and $V$ is the predicted azimuth $\theta$. 
For a given element $e_i$, $P(v_i|s_i)$ encodes the probability that the user correctly recognizes the element location at $v_i$ when the corresponding audio cue is played at $s_i$. 
Zero localization blur would lead to a perfect $P(v_i|s_i)$, represented as 1.0 along the diagonal \autoref{fig:confusion_matrix}.
Therefore, we use $P(V|S)$ to measure intensity of the localization blur for the element layout.

\subsubsection{Cone of Confusion}
The probability distributions only capture the probability under one-element conditions where $|E|=1$ because, in the data collection, only one audio cue was played at a time, and participants predicted one location for that cue.
When there are multiple elements in the layout, however, additional confusions need to be considered, such as whether two or more virtual elements lie on each other's cone of confusion (\autoref{fig:cone_of_confusion}).
\rv{\system{} calculates the cone of confusion by finding a circle perpendicular to the ground, or parallel to the head's vertical front-back plane, intersecting that element.}
The analyzer calculates the shortest spherical distance $D$ of the cone of confusion that a sound belongs to from the other elements' visual locations:
\begin{equation}
    D(v_i, s_i) = \min_{v_j \in V, v_j \neq v_i}{|c(s_i)-v_j|}  
\end{equation}
where $c(s) = \argmin_{v_c \in V_c}{|s - v_c|}$ and $V_c$ is the set of locations that are on the same cone as $s$. A larger distance $D(v_i, s_i)$ indicates a lower chance of cone-of-confusion errors.

\subsection{\system{} Optimizer}
The optimizer aims to find the optimal placement of spatial audio cues, denoted as $S^*=(s_1, \ldots, s_n)$, based on the corresponding localization blur probabilities $P(V|S^*)$ and cone of confusion distances $D(V, S^*)$.
It takes the visual locations $V$ as input, 
In our target scenarios, users should be able to accurately identify the correct visual location $v$ corresponding to each audio cue $s$, with no confusion from other potential perceived locations in $V$. 
We use integer programming to the find the optimal placement $S^*$ by weighing the data-based probability and possible effects of cone of confusion.



The optimization is performed over a discrete space divided into bins of a configurable bin size of 12\degree{} (\ie 30 horizontal bins total).
We chose this bin size based on the results from Valzolgher~\etal\cite{valzolgher2020impact}, who measured a typical azimuth error of 11.5\degree{}.
The integer program seeks to assign each visual location $v$ to a sound location $s$ such that the total likelihood is maximized, where $S_{all}$ denotes the set of all possible sound locations in bins ($S_{all} = \{12\cdot n\ |\ n\in[1,2,...,30]$):
\begin{equation}
\max \sum_{v \in V} \sum_{s \in S_{all}}  (w_{blur} \cdot P(v|s) + w_{cone} \cdot D(v, s)) \cdot x_{v, s}
\end{equation}

subject to
%
%
\begin{equation}
x_{v, s} \in \{0, 1\} \quad \forall v \in V, \forall s \in S_{all}
\end{equation}

Here, $x_{v, s}$ is a binary variable indicating whether visual location $v$ is assigned to sound location $s$.
The values for $w_{blur}$ and $w_{cone}$ were set to 0.9 and 0.1, \rv{respectively}, based on preliminary testing.

\subsubsection{Constraints}
We introduce a set of additional constraints in order to avoid duplicate elements, overlapping or flipped sound sources between different elements.
A flipping behavior refers to a change in the clockwise ordering of elements from 0\degree{} to 359\degree{}, \eg an element is left of another one before the optimization, it should stay on the same side after the optimization.

We ensure a bijective assignment that each visual element is only assigned to one bin by enforcing
\begin{equation}
    \sum_{s \in S_{all}} x_{v, s} = 1 \quad \forall v \in V
\end{equation}

In order to ensure that elements do not overlap or flip with each other we also keep a set of cumulative variables $y$, where 
\begin{equation}
    y_{v, s} = \sum_{j \in \{s+1, ...\}} x_{v, j} \quad  \forall v \in V
\end{equation}

Given that the visual elements were originally ordered, we then introduce the constraints
\begin{equation}
    y_{v, s} \leq y_{v-1, s-1} \quad  \forall v \in V, \forall s \in S_{all}    
\end{equation}

This ensures that elements maintain their relative ordering and also that they do not overlap with each other.

The output of the optimizer is a set of sound locations $S^*$, which are assigned to the corresponding visual locations ($V$) of the layout. 
It minimizes errors caused by both localization blur and cone of confusion as a whole. 
The location set is determined as follows:
\begin{equation}
    S^* = \{ s~|~\exists v \in V, x_{v, s} = 1\}    
\end{equation}


\subsection{Computational Considerations}
Our discrete optimization enables nearly real-time computation of optimal placement of spatial audio cues.
Tested on an Alienware x17 R2 laptop with a 12th Gen Intel Core i9 processor, our integer program finds an optimal sound locations $S^*$ in a runtime less than 10~ms for 2, 5, and 10 virtual elements.
The runtime on average of 5 runs was 34.8~ms, 101.6~ms, and 165.2~ms for 20, 50, and 100 elements, respectively.

\section{User Evaluation} \label{sec:evaluation}
We evaluated the effectiveness of \system{} in enabling users to disambiguate spatial audio cues in XR through a user study.

\subsection{Study Design}
Given a layout in XR, participants were tasked to identify which visual element was the source of audio notifications.

\autoref{fig:notification_assets} illustrates the visuals of the virtual elements and spectrograms of the audio cues used in the study.
\rv{We chose complex waveforms instead of a pure tone to reduce confounding factors in evaluating displacement effects as explained in Section~\ref{sec:background_localization}.}
We used a within-subject design with two independent variables, \textit{Method} and \textit{Visibility}. 
For method, we compare three different levels:
\begin{itemize}
    \item \textbf{Generic HRTF}: Baseline spatial audio where an audio cue is played at its visual location ($S$ = $V$).
    \item \textbf{Dynamic audio}: Dynamic audio cue moving in the direction towards its visual location~\cite{barde2016attention}.
    \item \textbf{\system{}}: Audio cues played at the sound locations ($S^*$) optimized by \system{}, which may be different from their visual locations ($V$). 
\end{itemize}

For Visibility, we tested two levels, specifically whether all elements are \textit{visible} or \textit{hidden}. For hidden, participants are made aware of the elements' positions in advance.
This condition was designed to simulate a future XR scenario where users place virtual elements in their surroundings so that they are not conspicuous to reduce clutter and occlusion~\cite{lindlbauer2019context, han2023blendmr}; or for audio-only XR approaches.
Three methods and two visibility levels led to six conditions in total. 
In each condition, participants completed five trials, \ie five target identification tasks.
Each condition was repeated twice.

\begin{figure}[t]
    \centering
    \includegraphics[width=\columnwidth]{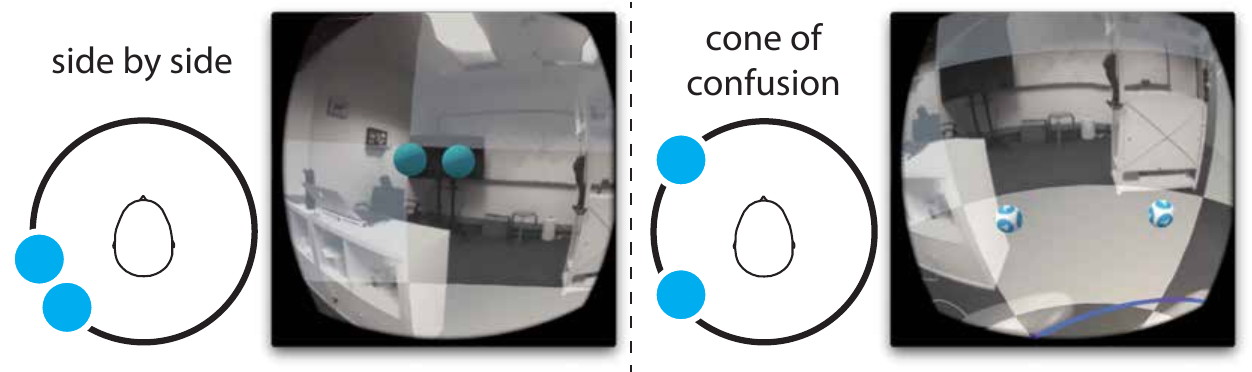}
    \caption{Examples of side-by-side and cone-of-confusion layouts in user study.}
    \Description{Examples of side-by-side and cone-of-confusion layouts in user study.}
    \label{fig:study_layouts}
\end{figure}

\paragraph{Layouts \& virtual elements.}
We included three types of layouts with different numbers of elements as random variables: side-by-side, cone of confusion, and random.
Side-by-side consisted of layouts where two virtual elements with the same application icon and notification sound were placed next to each other, separated by 12\degree{} apart in azimuth with the same elevation~(\autoref{fig:study_layouts} left).
For cone-of-confusion layouts, two similar virtual elements were placed on the front/back of each other at different angles ((\autoref{fig:study_layouts} right). 
The both layout types, the choice of application was randomized.
For random layouts, we placed five virtual elements randomly in space, while ensuring that at least two instances of the same application type were present to further test disambiguation.
As virtual elements, we used representations of one of five choices (Discord, Messenger, Slack, Snapchat, Telegram) with the accompanying sounds, assigned to conditions randomly.

\paragraph{Dependent variables.}
For dependent variables, we measured how often participants identified the correct virtual element as sound source, and the response time in milliseconds.

\begin{figure*}[t]
    \centering
    \includegraphics[width=\linewidth]{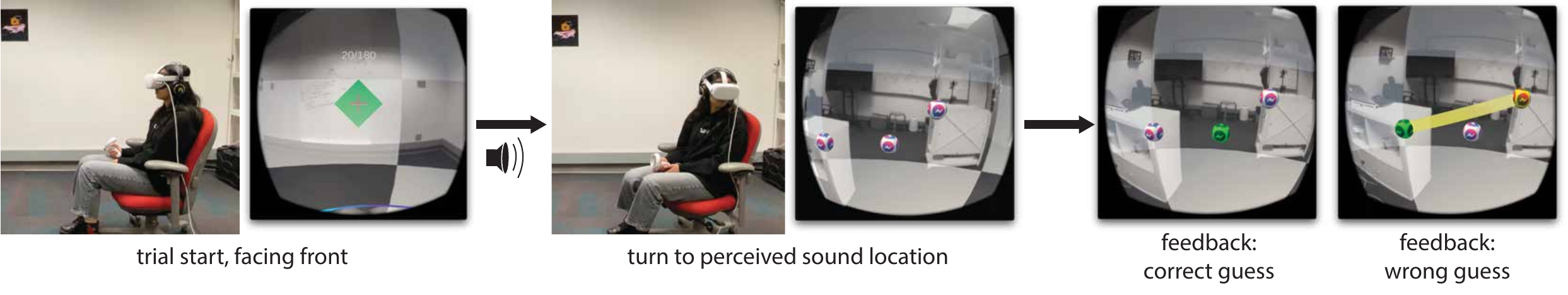}
    \vspace{-2.5em}
    \caption{Procedure of each trial in user evaluation. A participant first faces front to the green diamond, aligns the cross, and clicks the button on the right controller. Then, the prototype plays a sound from one of the elements. The participant turns towards a location they think the sound was played from, and clicks to confirm. The participant can review the correct answer in green and their selected response in yellow. This example shows an example of random layout.}
    \Description{Illustration of the procedure of each trial in user evaluation. A participant first faces front to the green diamond, aligns the cross, and clicks the button on the right controller. Then, the prototype plays a sound from one of the elements. The participant turns towards a location they think the sound was played from, and clicks to confirm. The participant can review the correct answer in green and their selected resopnse in yellow. This example shows an example of random layout.}
    \label{fig:evaluation_study_setup}
\end{figure*}

\subsection{Apparatus and Participants}
Participants wore a Meta Quest 2 headset and a pair of AKG Pro K240 Studio over-ear headphones. 
The study program was developed using Unity version 2021.3.9 and HRTF spatial audio from the Oculus Audio SDK version 47.0.
The virtual environment was similar to the data collection study.
In contrast to the data collection, this study was done in a Mixed Reality setup with virtual elements in a passthrough-enabled environment to resemble our target scenarios.
The setup is shown in \autoref{fig:evaluation_study_setup}.

We recruited 12 participants (7 female, 5 male) between 20 and 27 years old (mean = 22.3, SD = 2.46) from a local university. 
Participants had an average experience with Augmented Reality of 2.58 (SD = 1.24), with Virtual Reality of 3.08 (SD = 1.24), and with spatial audio of 2.75 (SD = 1.29), one a scale from 1 (None) to 5 (Expert).

\subsection{Procedure}
Participants were first given an introduction of the study, signed the consent form, and filled out the demographic questionnaire. 
The experimenter then ran a tutorial session where participants experienced the task using the generic HRTF. 
Participants also adjusted the volume to a comfortable level.
After the tutorial, participants proceeded to the main study which involved a total of 180 trials, \ie 30 trials per condition. 
The study was divided into two blocks of 90 trials with a break in between.
Each block consisted of six conditions (3 methods and 2 visibility levels) of which order was counterbalanced using a balanced Latin square. 
For each condition, three layout types appeared in a randomized order. 
For each layout type, the target element was randomly selected for each of 5 trials. 
The three layouts and their elements were kept consistent across conditions for a fair comparison. 
The set of layouts and elements differed across blocks. 

Before each trial, participants were instructed to look at each virtual element and confirm its position by pressing a controller button to gain familiarity with the current layout.
For this part, virtual elements were visible for both levels of \textit{visibility}.
Then, for each subsequent trial, participants faced forward, and started the trial by pressing a controller button. 
The spatial audio cue was played using the current condition, and participants rotated to the respective virtual element.
They then pressed a button to complete the trial.


\subsection{Results}
We analyzed the percentage of correct source identification and response time for different conditions.
We use these as proxy metrics of confusion, with higher accuracy indicating less confusion, and faster response time indicating higher confidence with less confusion.
In the analysis, we removed outliers detected using the IQR method based on the response time. 
For source identification accuracy, we used the general linear mixed effects regression analysis to extract method-accuracy and visibility-accuracy relationships by running \texttt{glmer} models using R and the lmerTest package. 
The analysis sets each method and visibility level as fixed effects and participant ID as a random effect. 
For response time, we used ANOVA and post hoc tests with Holm-Bonferroni correction for significance tests. 

\begin{figure*}[htbp]
    \centering
    \vspace{-0.5em}
    \includegraphics[width=\linewidth]{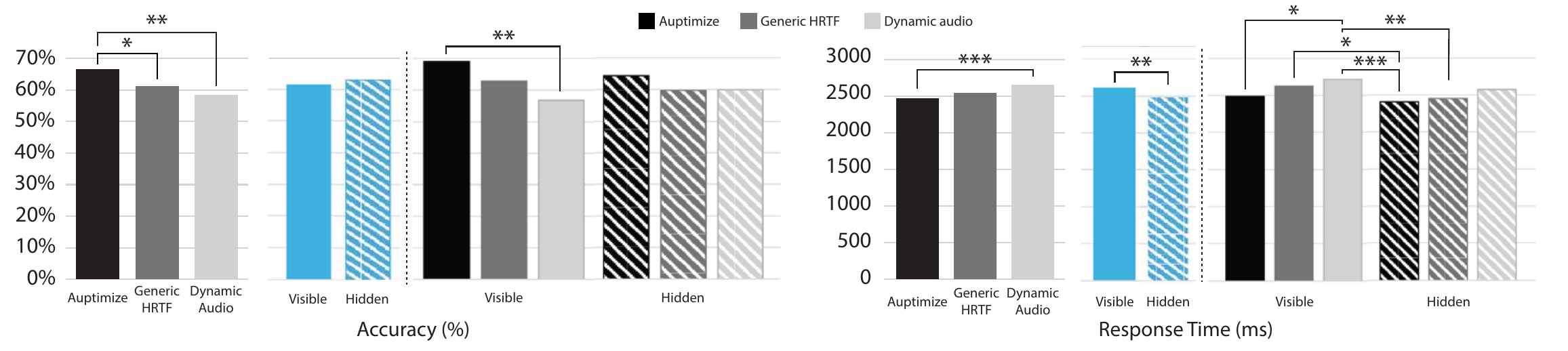}
    \vspace{-2em}
    \caption{Source identification accuracy (percentage of correct identification) and response time (in milliseconds) for different methods and visibility.}
    \Description{Box plots of source identification accuracy (percentage of correct identification) and response time (in milliseconds) for different methods and visibility. Auptimize leads to higher accuracy and faster response times.}
    \label{fig:results}
\end{figure*}

\autoref{fig:results} summarizes comparisons across different methods and two visibility levels.
\autoref{tab:results_layout_types} shows each method's performance for different layout types.

\begin{table}[htbp]
\centering
\caption{Summary of source identification accuracy and response time for different methods and layout types.}
\begin{tabular}{lcc}
\toprule
\textbf{Layout Type} & \textbf{Accuracy (\%)} & \textbf{Response Time (ms)} \\
\midrule
\rowcolor{gray!20} \textbf{\system{}} & 66.6\% & 2444.8 \\
Side by side & 71.3\% & 2493.2 \\
Cone of confusion & 67.3\% & 2218.6 \\
Random & 61.2\% & 2622.6 \\
\midrule
\rowcolor{gray!20} \textbf{Generic HRTF} & 61.3\% & 2538.4 \\
Side by side & 63.5\% & 2543.5 \\
Cone of confusion & 56.3\% & 2383.6 \\
Random & 63.9\% & 2684.1 \\
\midrule
\rowcolor{gray!20} \textbf{Dynamic Audio} & 58.4\% & 2644.7 \\
Side by side & 56.6\% & 2612.5 \\
Cone of confusion & 61.0\% & 2559.8 \\
Random & 57.7\% & 2758.4 \\
\midrule
\bottomrule
\vspace{-1em}
\end{tabular}
\label{tab:results_layout_types}
\end{table}

\subsubsection{Source Identification Accuracy}
Our analysis revealed that \system{} (66.6\% accuracy) significantly outperformed the generic HRTF method (61.3\% accuracy)  with a moderate effect size (estimate = $-0.24$, $SE$ = 0.12, $z$ = $-2.05$, $p$ <. 05).
Similarly, \system{} demonstrated a significantly higher accuracy compared to the dynamic audio method  (58.4\% accuracy) , with a stronger effect (estimate = $-0.36$, $SE$ = 0.12, $z$ = $-3.11$, $p$ <. 01).
These results were observed when the method was the only fixed effect in the model, as shown in \autoref{fig:results}.

When both method and visibility were included as fixed effects, the advantage of \system{} over generic HRTF was present, but the evidence was less compelling, indicating only a marginal improvement (estimate = $-0.27$, $SE$ = 0.17, $z$ = $-1.66$, $p$ = .09).
A more statistically significant improvement was observed against dynamic audio when virtual elements were visible (estimate = $-0.55$, $SE$ = 0.17, $z$ = $-3.28$, $p$ < 0.01), suggesting a substantial effect, as shown in \autoref{fig:results}. 
However, this improvement was not statistically significant in conditions where virtual elements were hidden from view.
As illustrated in \autoref{tab:results_layout_types}, \system{} improves source identification accuracy especially in side-by-side and cone-of-confusion layouts.

\subsubsection{Response Time}
The response time for auditory source identification using different methods and visibility are shown in \autoref{fig:results}.
Significant differences were observed between \system{} and dynamic audio ($M$ = $-200.862$, $SE$ = 49.279, t = $-4.076$, $p$ < .001, Cohen's d = $-0.222$).
The medium effect size suggests that \system{} significantly outperforms the dynamic audio method with a substantial mean decrease in response time. 
The comparisons between \system{} and generic HRTF ($p$ = 0.060), and between generic HRTF and dynamic audio ($p$ = 0.055) did not reach statistically significant difference.

Between visible and hidden conditions of virtual elements, a statistically significant decrease in response time was observed in the hidden condition ($M$ = $-130.399$, $SE$ = 40.138, $t$ = $-3.249$, $p$ < .01, Cohen's d = $-0.144$), indicating that response times were quicker when elements were not visible. 

In the interaction between method and visibility for response times, in the visible condition, \system{} demonstrated a significant reduction in response time compared to the dynamic audio ~($M$ = $-228.220$, $SE$ = 70.692, $t$ = $-3.228$, $p$ < .05, Cohen's d = $-0.252$), indicating a medium effect size. 
\system{} with hidden elements had a significant reduction compared to generic HRTF ($M$ = $-219.091$, $SE$ = 68.927, $t$ = -3.170,  $p$ < .05, Cohen's d = $-0.242$) with a small but noticeable effect size and dynamic audio ($M$ = $-311.456$, $SE$ = 69.774, $t$ = -4.464,  $p$ < .001, Cohen's d = $-0.344$) with a closer to medium effect size in the visible condition.
Generic HRTF with hidden elements also showed a significant reduction compared to dynamic audio with visible elements ($M$ = $-262.375$, $SE$ = 70.020, $t$ = $-3.747$, $p$ < .01, Cohen's d = $-0.290$).

\section{Example Applications}
We believe that \system{} is beneficial for a wide range of XR applications that rely on audio cues. 
In the following, we describe several specific examples highlighting its applicability.

\begin{figure}[h]
    \centering
    \includegraphics[width=\columnwidth]{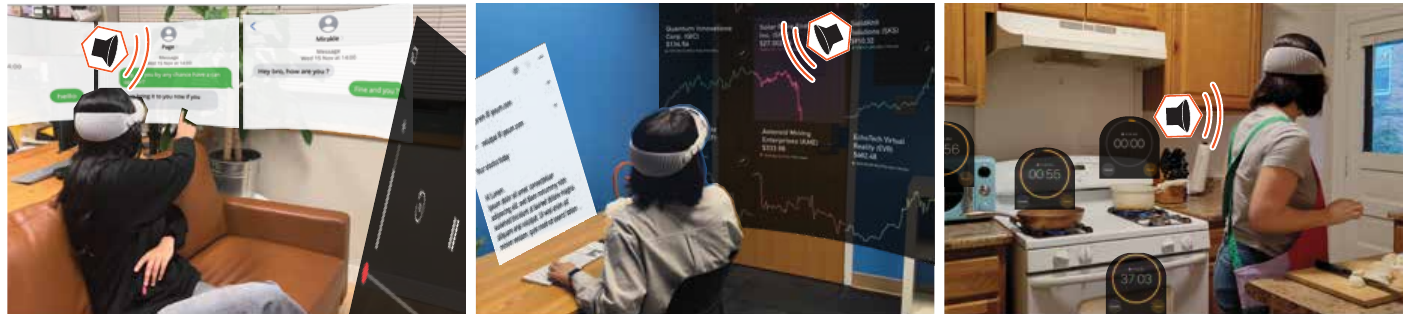}
    \caption{Example applications of spatial notification and alerts. \textit{Left}: user engaging in multi-window messaging. \textit{Center}: trader gets notified about a change in stock price. \textit{Right}: cooking application with spatialized audio timer. }    
    \Description{Example applications of spatial notification and alerts. Left: user engaging in multi-window messaging. Center: trader gets notified about a change in stock price. Right: cooking application with spatialized audio timer.}
    \label{fig:applications_notifications}
\end{figure}

\subsection{Spatial Notifications and Alerts}
\system{} can be integrated with existing notification or alert systems to provide users with better awareness of virtual and augmented physical objects~(\autoref{fig:applications_notifications}).

\paragraph{Multi-window messaging}
An XR user is involved in multiple simultaneous conversations in different messenger windows and a feed of stories.
Using conventional spatial audio, when they receive a notification, identifying the source would be challenging without deeper inspection.
Using \system{}, the user can more clearly identify the source window, and react faster without the additional mental effort during search.


\paragraph{Stock trading application}
A stock trader uses a very large XR display with dozens of windows all around them, each corresponding to a different stock.
Often, these windows trigger an audio notification when stock prices change, and the user will need to pay close attention to this fluctuation. \system{} optimizes the location from which this sound plays to more accurately direct the user's attention toward the window of interest, enabling them to react.

\paragraph{Cooking}
\system{} also be used to improve awareness of augmented physical objects. 
For example, a user may perform many simultaneous tasks while cooking in a kitchen, such as chopping vegetables while boiling pasta and baking.
This requires them to keep track of when individual tasks are finished. 
\system{} can better spatialize the auditory timer notifications to improve the user's localization of where the sound is coming from. 

\begin{figure}[h]
    \centering
    \includegraphics[width=\columnwidth]{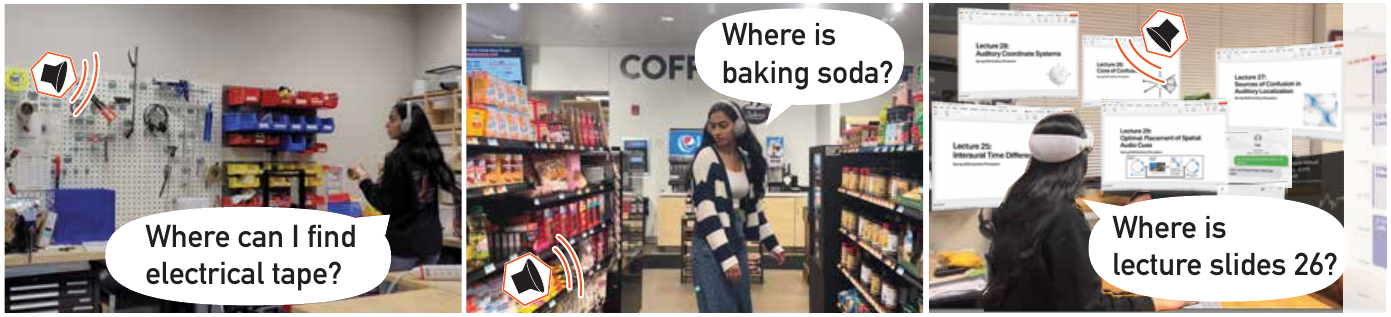}
    \caption{Example applications of audio-based interactive guidance, including a search for physical tools (left), grocery items (middle), lecture contents (right).}
    \Description{Illustration of example applications of audio-based interactive guidance, including a search for physical tools (left), grocery items (middle), lecture contents (right).}
    \label{fig:applications_interactive}
\end{figure}

\subsection{Audio-based Interactive Guidance}
\system{} can be used as an additional form of audio-based guidance for interactive localization tasks for both virtual objects and with augmented physical objects, as detailed in \autoref{fig:applications_interactive}. Spatial audio cues can be extremely helpful for interaction with difficult-to-locate and out-of-view objects where visual cues are insufficient. The addition of an auditory cue can help users better understand the position of an object in a scene.


\paragraph{Audio-only XR - Search}
\system{} help in audio-only XR scenarios. 
In a cluttered machine shop, for example, audio cues can help users locate objects they are searching for.
In this case, the elements the our system considers are all of the physical items around the user, allowing \system{} to produce a sound that best helps the user to localize the position of their search target.
This method is most effective if information about items is available, \eg stocking information for store shelves.
Similar guidance can be applied to different environments, such as supermarkets, workshops, or libraries.


\paragraph{Audio-only XR - Navigation}
Navigation is another common scenario that this audio-only AR implementation can be applied to. 
For example, if a user wants to find a store on a crowded street or within a mall, an audio cue from \system{} can be used to direct the user's attention towards that store.

\section{Discussion}
Our user evaluation shows that \system{} effectively enhances ambiguous audio source identification accuracy and reduces response time in comparison to generic HRTF and dynamic audio. 
This evaluation result along with the collected data (Section~\ref{sec:optimal_sound_displacement}) highlights that the traditional notion of placing the sound source of a virtual element at the same location as its visual part is not the best choice for accurate localization and identification. 
\system{} provides better locations for audio cues by disentangling them from the visual locations.

Despite prior work suggesting that a dynamic audio cue reduces the onset time for target acquisition~\cite{barde2016attention}, our study showed that a dynamic audio cue lowers the accuracy of sound source identification with longer search time. 
This difference is mainly attributed to different experiment settings, which focused on a setting where the target is not ambiguous ($\pm$50\degree and $\pm$100\degree). 

\subsection{Relationship between \system{} and HRTF}
Existing work focuses on ways to improve personalized HRTF modeling in order to enhance auditory localization. 
\system{} is a complementary adjustment on top of HRTFs and addresses the problem of localization-based audio source disambiguation, which also exists in the physical world in human's natural auditory system. 
We leverage the ventriloquist effect to move the locations of audio cue sources to the optimal locations that are most distinct from other sources, while remaining identifiable.

\subsection{\rv{Generalization and Personalization}}

\rv{Sound localization errors may vary across different hardware, HRTFs, and users. 
In this work, we used the state-of-the-art HRTF at the test time (Oculus Audio SDK 47.0) and stereo headphones (AKG K240 over-ear, semi-open headphones) similar to audio perception studies. 
We expect that other headsets will have similar results with stereo headphones and the same HRTF.
However, further evaluation in diverse setups is needed to test the scalability of our approach.
}

\rv{In addition, we opted to build a generic model by using the aggregated probability $P(V|S)$ instead of personal calibration.
Our aggregated model improved target identification accuracy for new participants in the evaluation.
This resonates with Berger et al.~\cite{berger2018generic} that non-individualized HRTFs are sufficient for auditory localization. 
However, there is still room for improvement in performance.
Applying personalization is an important avenue for future investigation, as using a personal probability instead of the aggregated one, or applying calibration, could further improve the performance of \system{}.
}

\subsection{Localization Data and Optimization}
We opted for discrete optimization instead of continuous optimization for two reasons.
First, our data is sparse and does not cover every angle, which makes it challenging to derive closed-form solutions.
Secondly, since we leverage the ventriloquist effect, which is effective within 30\degree{} of azimuth angle, the optimization based on the bin size of 12\degree{} is still sufficient to relate the audio cue to its visual host. 
We hope to explore larger datasets and other optimization methods for more accurate modeling of auditory localization behaviors in XR in the future.

Our data was collected for a hidden single sound that could come from any direction. 
This is different from our target scenarios with multiple virtual elements.
While our approach showed to be effective, we believe that data of multiple visible elements might enable us to model localization blur with even higher accuracy.


\subsection{Complex Auditory Scenes}
The current \system{} system operates under a set of assumptions about auditory settings that might not scale to all scenarios of XR usage.
For example, the distance of all sound cues in the studies were fixed at the sphere radius of 1.5 meters.
Localization of cues at different distances depends on the relative amounts of direct and reverberant energy, which relies on the understanding of the room structures that affect the echoes.
We hope to incorporate such parameters in our model in the future.

The studies were conducted in a quiet room without environmental noise.
In the real-world XR usage scenarios, however, varying setups of user's auditory environment would require modifications to the proposed spatial audio cue placement optimization. 
For example, if an environmental noise is in the same direction of a virtual element's sound source, the noise may mask the audio cue and hinder hearing of this cue. 
As a future work, the phenomenon of binaural unmasking~\cite{hirsh1948influence} can be utilized, where the detection of the audio signal can be improved by shifting the phase of the signal differently to both ears, or duplicating the directional noise to the opposite ear to unmask the masking effect of noise.

Furthermore, in the studies, sounds were played sequentially, only one sound being played at a time. 
When two sounds are present at the same time, binaural interference occurs, including across-frequency integration of binaural bandwidth and binaural beats.
Future work could study an advanced method to optimize the delivery of spatial audio cues by considering binaural interference of multiple sounds at a time. 
Besides the placement of audio cues, future systems could incorporate modulations of other audio properties such as frequency, phase, beats, and temporal envelopes.

\subsection{Audio-visual Perception}
Considering the effects of audio-visual perception, more factors beyond than auditory perceptual errors can be integrated in future work. 
For example, the ventriloquist effect becomes stronger when the visual element moves with a similar frequency or pattern that matches the auditory stimuli. 
Beyond static visual appearance of virtual elements explored in our study, future work could study and integrated the effects of animations of motions of virtual elements. 
Also, XR users leverage semantic connections between physical objects in the environment and virtual interfaces for placement~\cite{cheng2021semanticadapt}. 
Whether this bias of semantic connection asserts an effect on auditory perception between audio cues and virtual elements or augmented physical objects (\eg bird sounds from the direction of the window vs. from a stove top) needs further investigation. 

\section{Conclusion}
We propose \system{}, an optimization-based approach to mitigate confusion in localization-based source identification in XR, such as localization blur and cone of confusion.
The optimal placement for spatial audio cues generated by \system{} shows significant improvement in source identification accuracy of spatial audio cues in XR and reduction in response time, compared to generic HRTF and dynamic audio methods.
Our finding emphasizes that decoupling the location of spatial audio cues from their visual host can lead to improved performance of identification among multiple sources.
We see \system{} as a key factor for improving the placement of sounds to decrease perceptual errors.
By making sound localization and disambiguation more accurate, we believe that future XR applications that leverage the whole space around users for interaction can become even more beneficial.

\begin{acks}
We thank Prof. Laurie Heller for her Auditory Perception course at Carnegie Mellon University and Yi Fei Cheng for his valuable feedback and suggestions.
\end{acks}


\bibliographystyle{ACM-Reference-Format}

\bibliography{ref}

\appendix\onecolumn
\section{Appendix}
\begin{figure*}[ht]
    \centering
    \vspace{2em}
    \includegraphics[width=\linewidth]{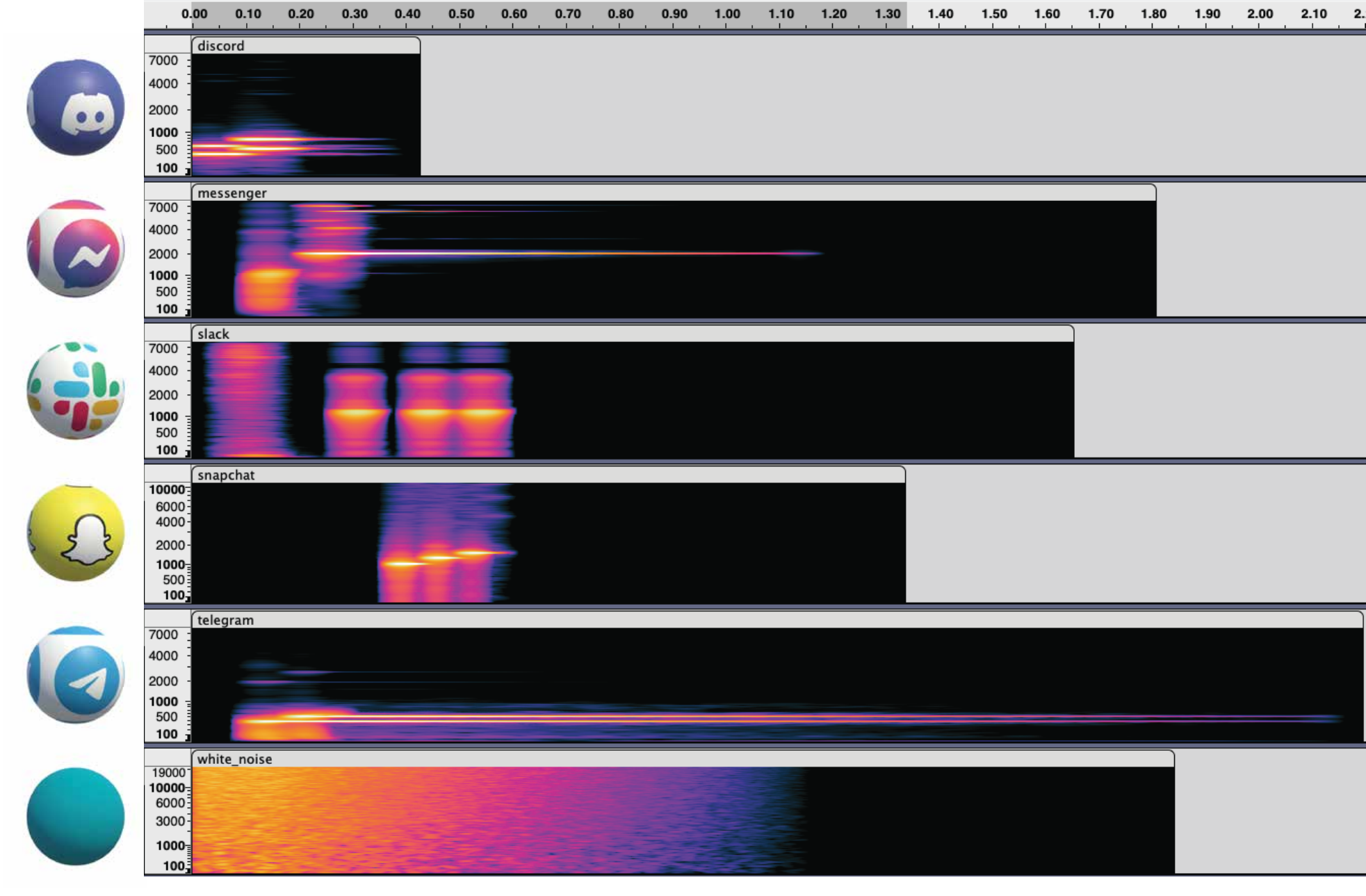}
    \captionof{figure}{Visuals of the messaging application icons and spectrograms of the corresponding message notification sounds used in the user evaluation study.}
    \Description{Visuals of the messaging application icons and spectrograms of the corresponding message notification sounds used in the user evaluation study.}
    \label{fig:notification_assets}
\end{figure*}

\end{document}